\documentclass{elsart}
\usepackage{epsfig}
\usepackage{lineno}

\begin{document}
\runauthor{Marco Battaglia}
\begin{frontmatter}
\title{Characterisation of a Pixel Sensor in 0.20~$\mu$m 
SOI Technology for Charged Particle Tracking}
\author[UCSC,LBNL,CERN]{Marco Battaglia,}
\author[INFN]{Dario Bisello,}
\author[LBNL]{Devis Contarato,}
\author[LBNL]{Peter Denes,}
\author[UCSC,INFN]{Piero Giubilato,}
\author[INFN]{Serena Mattiazzo,}
\author[INFN]{Devis Pantano,}
\author[UCSC]{Sarah Zalusky}
\address[UCSC]{Santa Cruz Institute of Particle Physics, University of California 
at Santa Cruz, CA 95064, USA}
\address[LBNL]{Lawrence Berkeley National Laboratory, 
Berkeley, CA 94720, USA}
\address[CERN]{CERN, Geneva, Switzerland}
\address[INFN]{Dipartimento di Fisica, Universit\'a di Padova and INFN,
Sezione di Padova, I-35131 Padova, Italy}

\begin{abstract}
This paper presents the results of the characterisation of a pixel sensor manufactured 
OKI 0.2$~\mu$m SOI technology integrated on a high-resistivity substrate, and 
featuring several pixel cell layouts for charge collection optimisation. The sensor 
is tested with short IR laser pulses, X-rays and 200~GeV pions. We report results 
on charge collection, particle detection efficiency and single point resolution.
\end{abstract}
\end{frontmatter}

\typeout{SET RUN AUTHOR to \@runauthor}


\section{Introduction}

Monolithic Si pixel sensors have become an established technology for precision 
vertex tracking in particle physics and in charged particle detection and photon imaging.
Commercial bulk CMOS is the most commonly adopted process for monolithic active pixel 
sensors (MAPS)~\cite{aps}. MAPS are being successfully applied to particle physics 
experiments~\cite{star,ilc}, imaging in transmission electron microscopy~\cite{emicro, 
deptuch, team} and photon imaging~\cite{barbier}. However, MAPS in bulk 
CMOS technology have several limitations. First, only nMOS or pMOS transistors can be 
built without disturbing the signal charge collection. 
Then, the charge collected from the thin epitaxial layer is small, limiting the achievable 
signal-to-noise ratio. The collection occurs through diffusion in an almost field-free 
region, which results in a long charge collection time and, possibly, increased sensitivity 
to radiation damage. The silicon on insulator (SOI) technology makes it possible to fabricate 
CMOS circuits on a thin Si layer, insulated from the handle wafer by a buried oxide layer 
(BOX). A high-resistivity Si handle wafer provides a sensitive volume that can be biased, 
thus improving both the speed of the charge carriers collection and the quantity of charge 
carriers generated by an ionising particle. Vias through the BOX connect the substrate to 
the electronics layer, where both nMOS and pMOS transistors can be built.

The appealing features of the SOI technology for developing pixel detectors were recognised a 
decade ago in a pioneering study~\cite{soi-iet1,soi-iet2,soi-iet3}. It has been with the 
availability of a commercial SOI process with small feature size CMOS from the collaboration 
between KEK, Tsukuba and OKI Industries Inc.\, that a systematic R\&D effort on SOI pixel detectors 
has started~\cite{soi-oki1,soi-oki2}. 
The KEK-OKI process has special vias cut into the BOX to contact the underlying handle wafer and 
collect the charge signal. 
A technology  proof of principle for charged particle tracking was obtained with the successful test 
of a prototype in OKI 0.15~$\mu$m SOI technology on a particle beam~\cite{Battaglia:2007eq}. 
Together with the detection of the first particles in the SOI pixels, one of the main challenges for 
the further development of the technology appeared. The bias applied to the high resistivity substrate 
induces a potential below the CMOS electronics layer which is shielded only in part by the BOX. 
This potential acts as a back-gate to the transistors shifting their thresholds. As a result of this 
``back-gating'' effect, the charge sensing and read-out electronics could only be operated for low to 
moderate depletion voltages $V_{d}$, up to 15-20~V, and thus the charge was collected by a depleted region 
of modest thickness. In that prototype sensor a series of guard-rings was used to counter the back-gating 
effect. Guard-rings take considerable space in the pixel and are only partially effective. 
Instead, the implant of a buried $p$-well (BPW) beneath the BOX and the transistors has been found to 
successfully screen the potential applied to the high resistivity layer in single transistor test 
structures~\cite{soi2}. This motivates the adoption of a BPW in the design of pixel cells immune 
from back-gating.

In this paper we present the characterisation of a prototype pixel chip fabricated in OKI 0.2~$\mu$m 
SOI technology, which implements different pixel cells designed to minimise the back-gating effect.

\section{Chip Description}

The prototype chip, named ``SOImager-2'' derives its global architecture from the earlier ``SOImager-1'' 
chip~\cite{pixel2010}. In order to optimise the design of the pixel cell for mitigating the back-gating 
effect and enhancing the charge collection, pixel cells of different design are implemented.
The chip is manufactured by OKI Industries Inc.\ in 0.2~$\mu$m SOI technology on $n$-type SOI 
wafers with a nominal resistivity of the handle wafer of  700~$\Omega\cdot$cm.

The sensitive area is a 3.5$\times$3.5~mm$^2$ matrix of 256$\times$256 pixels arrayed on a 13.75~$\mu$m 
pitch. In order to increase the speed of the serial read-out, the pixel matrix is divided into four 
parallel arrays of 64 columns each. Each array is connected to four identical parallel output analog 
stages. Both the pixel cell and the ancillary electronics are designed and tested to operate at up to 
50~MHz read-out frequency. The chip read-out implements a global shutter. Guard-rings are implemented 
around the pixel matrix ($p+$ I/O) and the peripheral I/O electronics ($p+$ Outer) in order to insulate 
the transistors from the effect of the potential below the BOX. 
The pixel cell keeps the same 3T design of the earlier prototype for the whole matrix, but the pixel array 
is now divided into eight different sectors, each implementing a different layout. These layouts explore 
different diode size, configuration of floating and grounded guard-rings around the pixel, as well as the 
use of a buried p-well, BPW. In this paper we report results for two pixel cell designs with no $p$-type 
guard-ring and BPW connected to the pixel diode. One pixel cell has a 1.5~$\mu$m diode with a large 
BPW layer extending just beneath the transistors. The second design features a larger diode (5~$\mu$m) 
with a BPW layer which extends below the diode only.

\section{Chip Tests}

$I-V$ and $C-V$ characteristics are obtained in the laboratory to determine the evolution of the 
leakage current with the depletion voltage, the breakdown voltage and to estimate the thickness 
of the depleted region. The sensor is then tested in the laboratory with short IR laser pulses 
of various wave-lengths and X-rays.
The laser diode is driven by a fast voltage pulse applied through a bias-tee which feeds the laser 
with a constant current. A 2~ns-long laser pulse is generated, transported using a single-mode optical 
fibre and focused to a $\simeq$~5~$\mu$m spot on the pixel front surface. The detector is mounted on 
a remotely controlled precision stage, which allows us to vary its position with an accuracy of 
$\le$~1~$\mu$m. The laser pulse is triggered in phase with the detector data acquisition cycle and has 
an adjustable delay. Laser tests allow us to measure the charge carrier signal and collection time as 
a function of $V_d$. By varying the laser spot position along pixel rows and columns, the charge sharing 
on neighbouring pixels is also investigated. X-rays are used to calibrate the pixel and study the 
linearity of its response with deposited charge. We use 5.9~keV X-rays from $^{55}$Fe sources and 
8.04~keV fluorescence radiation from a Cu target illuminated with a monochromatic synchrotron radiation 
beam at the LBNL Advanced Light Source (ALS).

\begin{figure}
\begin{center}
\begin{tabular}{cc}
\epsfig{file=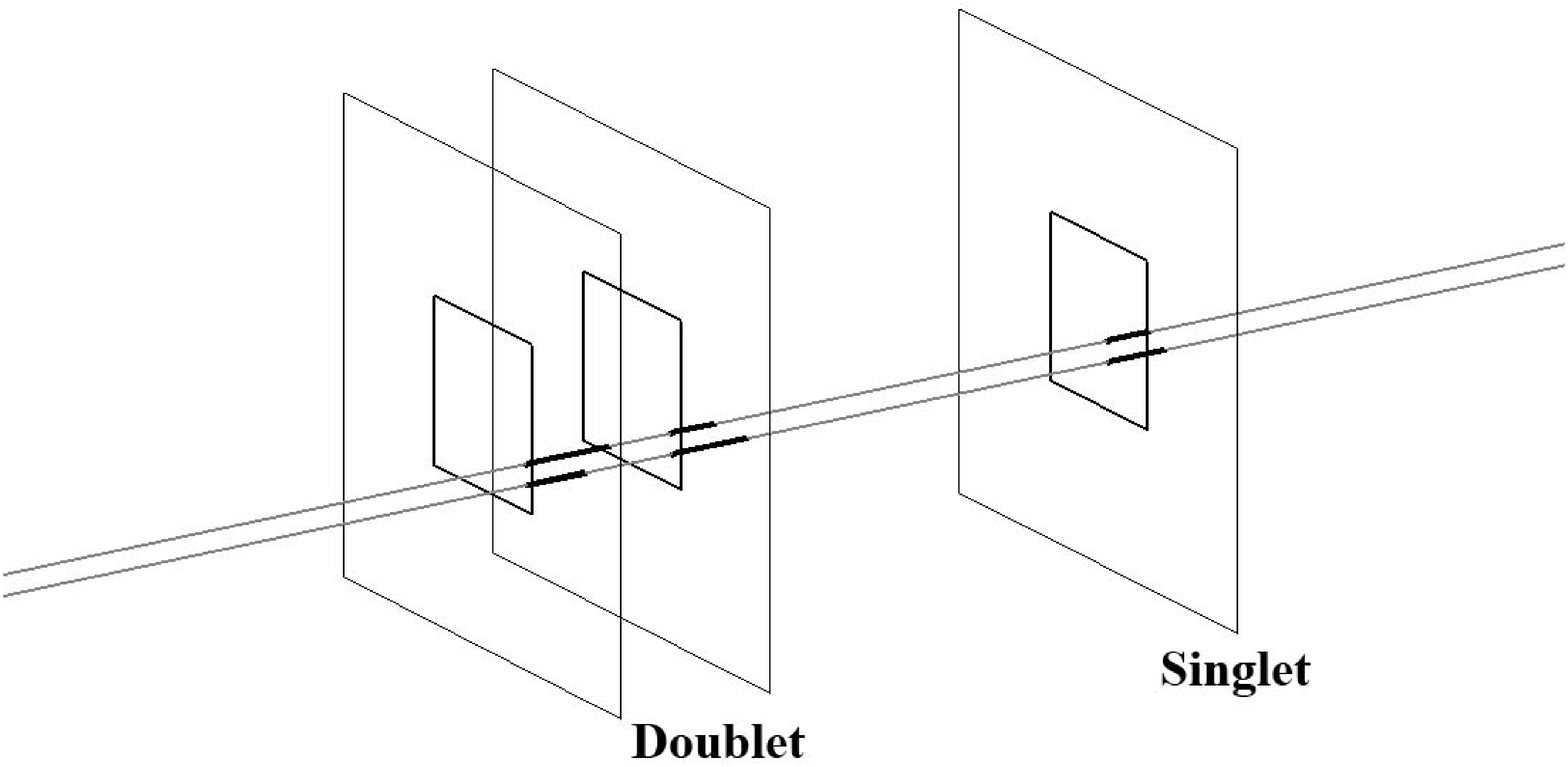, height=4.75cm} & 
\epsfig{file=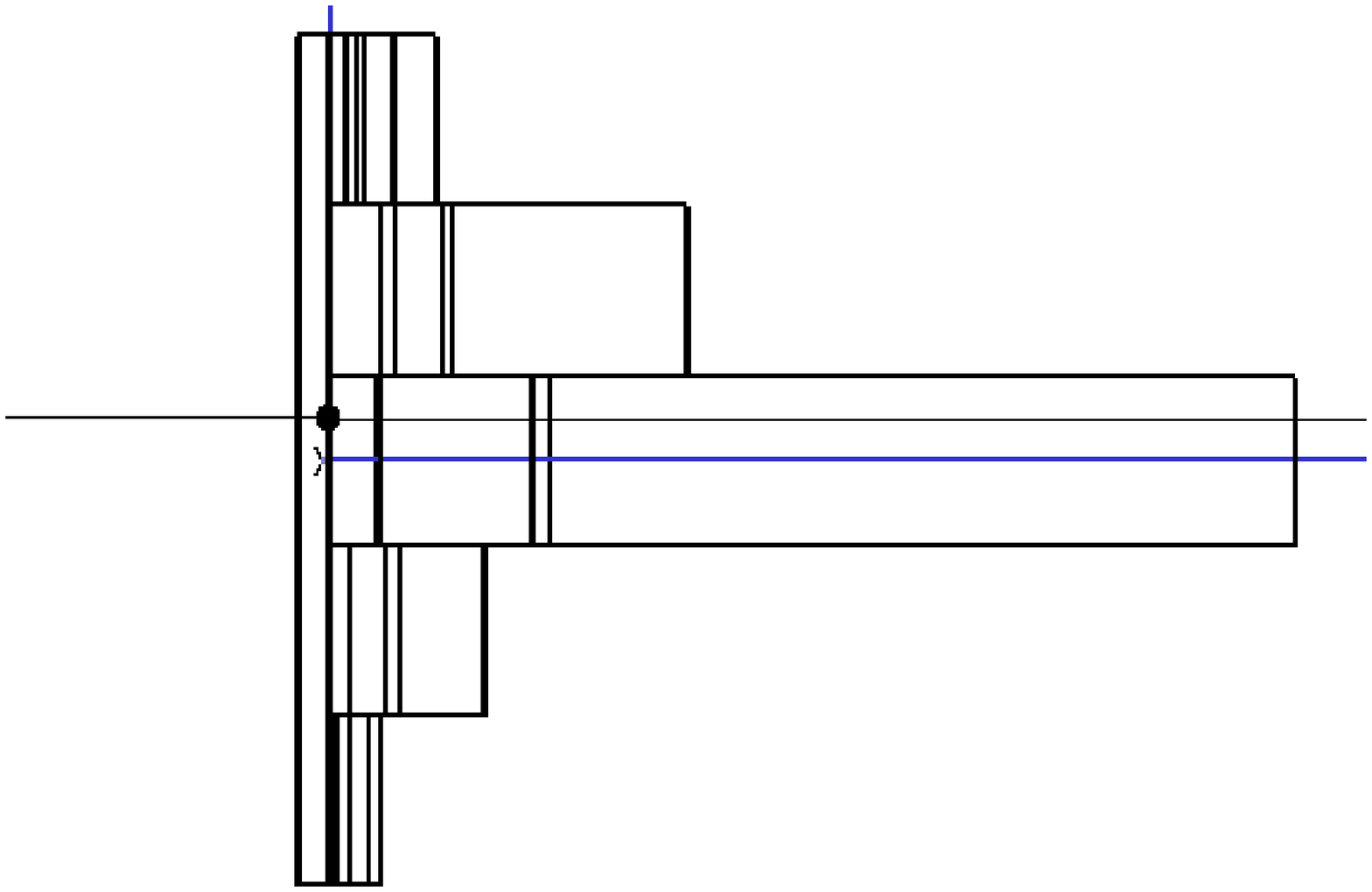, height=3.5cm} \\
\end{tabular}
\end{center}
\caption{(Left) Schematic layout of the SOI sensors for the beam test. The first two planes on the 
left make up the detector doublet, the plane on the right, called singlet, was mounted on a remotely 
controlled rotation stage. Two pion tracks reconstructed in a single event are shown. (Right) 
Individual pixel response in a cluster on the second doublet plane associated to an extrapolated 
track.} 
\label{fig:evt}
\end{figure}

The response to energetic charged particles is studied with 200~GeV/$c$ $\pi^-$ on the H4 
beam-line at the CERN SPS. Three SOImager-2 sensors have been arranged in a beam telescope 
configuration (see Figure~\ref{fig:evt}). The first two chips, spaced by 9.4~mm, are mechanically 
aligned and surveyed using a metrology machine before installation into a doublet unit. The third 
chip, located 36~mm downstream (called singlet in the following), is installed on a remotely controlled 
rotation stage. After installation on the beam-line, the three sensors have been aligned by optical survey. 
The final alignment is performed using particle tracks. Data is acquired at a rate of $\sim$150 events 
s$^{-1}$ during the 9.8~s-long SPS extraction spill. The detector temperature is kept stable at 
(20$\pm$1)~$^{\circ}$C by flowing cold air inside the optical enclosure housing the detectors.

For the lab and beam tests, the chip is mounted on a mezzanine board, which carries four analog pre-amplifiers, 
each receiving the signal from one of the four chip analog outputs. Signals are routed into differential lines 
and the signals are fed to the data acquisition system through 1~m-long twisted pair cables with standard USB 
connectors. The data acquisition system consists of an analog board pigtailed to a commercial FPGA development 
board, used as control unit~\cite{Battaglia:2009daq}. Both boards are powered by a stand-alone integrated power 
supply, which also supplies the detector bias. The analog board has five independent analog differential 
inputs, four of which have been used to read-out one analog output from each of the detectors in the 
doublet and two outputs of the singlet. Each differential input stage feeds a 100~MS/s 14-bit ADC. 
The control board has a Xilinx Virtex-5 FPGA which supplies the clocks and the slow control signals 
driving the sensor chips and routes the digitised data to a high speed FIFO to be formatted and 
transferred to the DAQ computer via a USB-2.0 link at a rate of 25~Mbytes/s. 
Most of the measurements have been performed with the chip clocked at 12.5~MHz, corresponding to
80~ns read-out time per pixel. The noise of the preamplifier stage and the read-out chain is 1.8~ADC 
counts. Data sparsification is performed on-line in the DAQ PC using a custom 
{\tt Root}-based~\cite{Brun:1997pa} program. Sensors are scanned for seed pixels with signal exceeding 
a preset threshold in noise units. For each seed, the 7$\times$7 pixel matrix centred around the seed
position is selected and stored on file. During the SPS beam test the pixel pedestal and noise 
values are updated at the end of each spill in order to follow possible drifts in their baselines.
Data are stored in {\tt Root} format and subsequently converted into {\tt lcio} format~\cite{Gaede:2005zz} 
for offline analysis. The data analysis is based on a set of custom processors in the {\tt Marlin} 
reconstruction framework~\cite{Gaede:2006pj}, which perform cluster centre-of-gravity reconstruction 
and cluster shape analysis, as well as track fitting for the beam test~\cite{Battaglia:2008nj}.

The final analysis of the sparsified data is performed offline. Clusters are reconstructed applying 
a double threshold method on the matrix of pixels selected around a candidate cluster seed. Clusters 
are requested to have a seed pixel with a signal-to-noise ratio, S/N, of at least 5.0 and the neighbouring 
pixels with a S/N in excess of 3.0. Clusters consisting of a single pixel are discarded. The cluster 
position is calculated as the centre of gravity of the measured pulse height. In the test with high 
energy pions, particle tracks are reconstructed using a straight line model, since the multiple 
scattering can be neglected at 200~GeV. 
We request the track slope in both coordinates to be smaller than 3$\times$10$^{-3}$. 
When we use all the three planes to reconstruct the tracks we also require the fit $\chi^2$ 
to be below 5. When we study the efficiency and the single point resolution we use only pairs of hits 
on the two layers not used as detector under test (DUT). Given the low particle density and 
the high read-out speed there are on average 0.86 hits/layer in non-empty events. This greatly 
simplifies the pattern recognition. The average reconstructed track multiplicity in non-empty 
events is 0.66. An example event display with two reconstructed particle tracks and the response of a 
cluster of pixels to a single pion hit are shown in Figure~\ref{fig:evt}.

\section{Detector calibration and charge collection studies}

\subsection{Pixel calibration and noise}

\begin{figure}
\begin{center}
\begin{tabular}{cc}
\epsfig{file=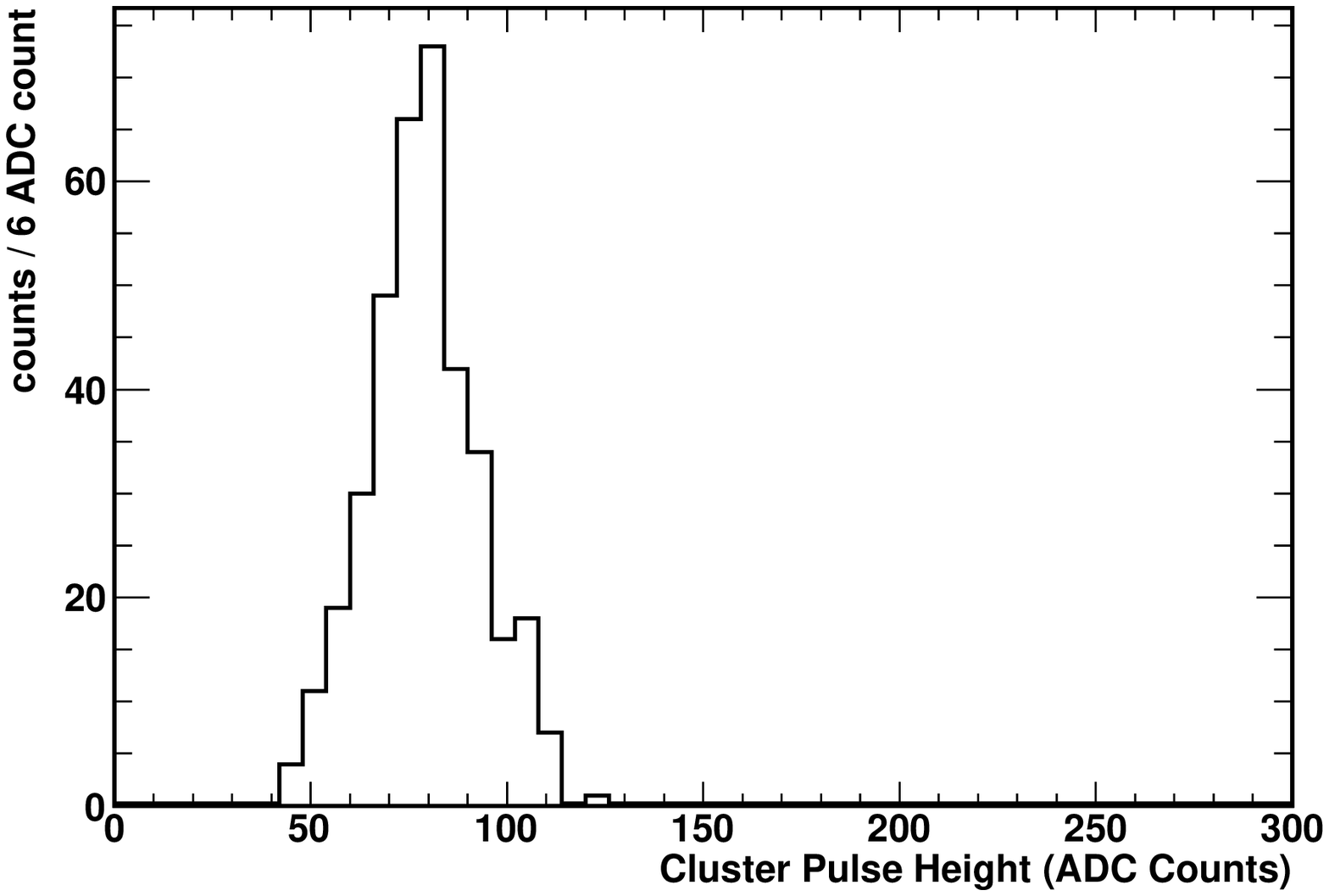,width=7.0cm} &
\epsfig{file=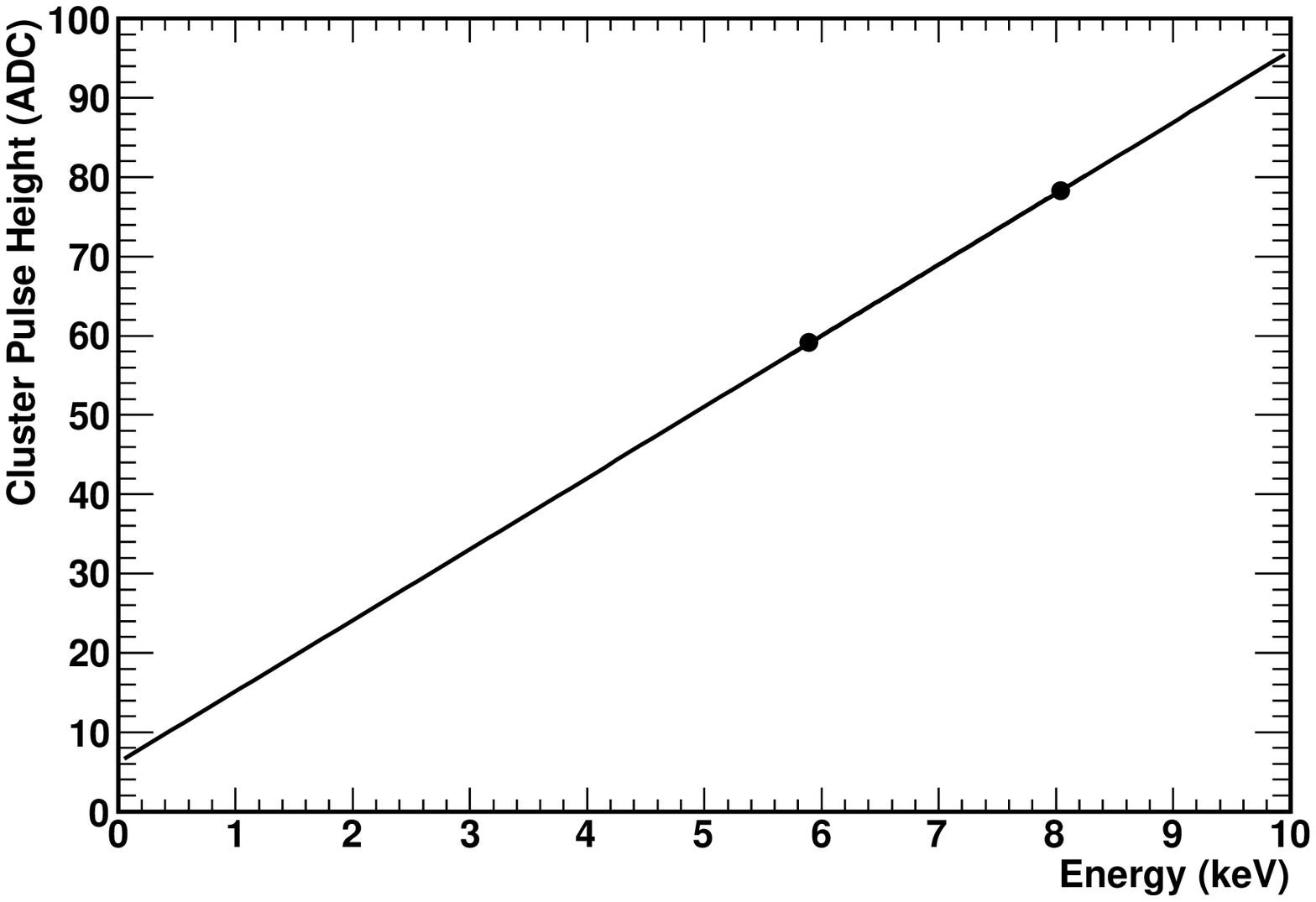,width=7.0cm} \\
\end{tabular} 
\end{center}
\caption{Response to X-rays: (left) cluster pulse height for 8.04~keV X-rays from Cu fluorescence 
and (right) most probable cluster pulse height values as a function of the X-ray energy for  
$^{55}$Fe and Cu fluorescence at $V_d$ = 50~V. The fitted slope corresponds to 
(30.3$\pm$1.8)~$e^-$ ADC 
count$^{-1}$.} 
\label{fig:xray}
\end{figure}

The ADC counts-to-electrons conversion is determined using the 5.9~keV X-rays emitted by a 
$^{55}$Fe source, which generate 1640 e$^-$s in the sensor substrate. Pixels with floating $p$-type 
guard-ring can be operated properly only up to $V_d$ = 20~V, since for higher voltages the 
back-gating effect sets in. This agrees with what already observed on previous sensor prototypes. 
Pixel cell designs with no $p$-type guard-ring and BPW connected to the pixel diode show 
the best performances and are used in the rest of this study. These sectors are fully 
functional up to $V_d$ = 100~V and the signal for 5.9~keV X-ray conversion is clearly visible 
in the cluster pulse height spectrum. The pixel gain and noise are studied as a function of the 
depletion voltage for various clock frequencies and operating temperatures.
We observe a decrease of the gain with increasing 
depletion voltage up to $\sim$50~V, when the gain becomes stable at 27.5~$e^-$ ADC count$^{-1}$. 
The pixel cells with BPW below the transistors and connected to the diode exhibit full charge 
collection in the diode but $\simeq$9\% lower gain compared to those with the BPW layer only 
connected to the diode, possibly due to the increased cell capacitance. The pixel calibration 
is verified using 8.04~keV Cu-fluorescence X-rays from the data collected in a run at the ALS. 
Results are shown in Figure~\ref{fig:xray}, which exhibits a linear scaling of the cluster pulse 
height with the X-ray energy.

\begin{figure}
\begin{center}
\epsfig{file=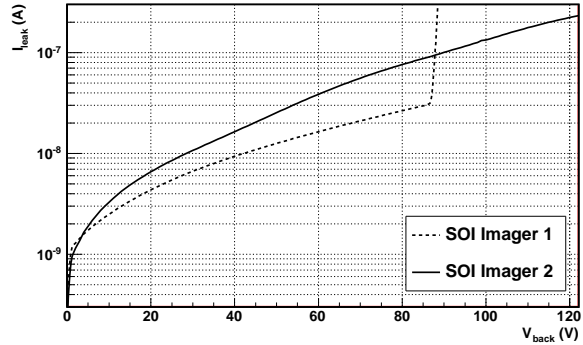,width=8.5cm} 
\end{center}
\caption{I-V curve measured for the SOImager-2 (continuous line) and the earlier SOImager-1 
(dashed line) sensors. The breakdown voltage of the two sensors is 140~V and 90~V, respectively.} 
\label{fig:leak}
\end{figure}

A set of $I-V$ measurements to determine the current flowing in the detector substrate 
as a function of the depletion voltage $V_d$ are performed by biasing the chip and 
monitoring the current with a DC source/monitor unit. In this measurement the depletion 
voltage is applied to the probe station plate. Two probes are used to measure the currents 
in the chip. The first measures the current to the pixel guard-ring grid and the other that 
to the $p+$ I/O implant, with the external guard-ring structure ($p+$ Outer) kept floating. 
The result is shown in Figure~\ref{fig:leak} with a comparison to the earlier sensor. The measured 
leakage current is of the order of 10-100~nA for 30 $< V_d < 90$~V. This result is in agreement
with the calculation of the leakage current from the increase of the pixel dark levels for 
different integration times. 
The pixel noise is due in part to leakage current in the substrate. The SOImager-2 sensor shows 
a larger leakage current compared to the previous SOImager-1 prototype produced in the same 
process in 2008 (see Figure~\ref{fig:leak}). 
The total single-pixel noise for 12.5~MHz read-out, measured in the lab and during the beam test, 
is (83$\pm$8)~$e^-$ ENC at room temperature. After subtracting the read-out noise in quadrature, 
we obtain the detector single-pixel noise of (73$\pm$7)~$e^-$ ENC  at $V_d$ = 50~V and room 
temperature. It decreases to (43$\pm$4)~$e^-$ ENC operating the sensor at -20$^{\circ}$C
and it remains stable when varying the pixel clock frequency, $f$, in the range 3$<f<$12.5~MHz.

\subsection{Depletion depth, charge carrier signal and collection time}

\begin{figure}
\begin{center}
\begin{tabular}{c}
\epsfig{file=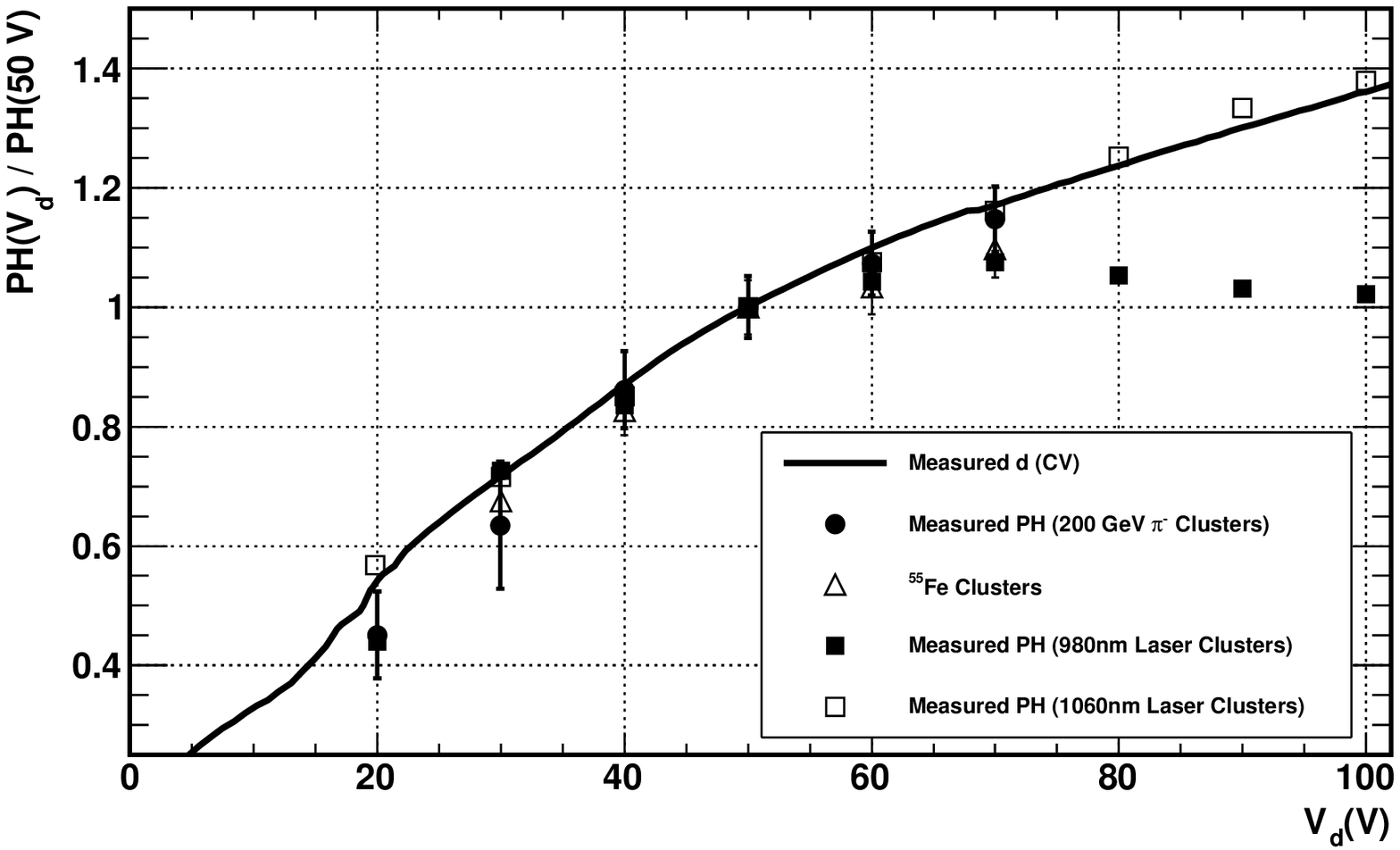,width=9.5cm} \\ 
\epsfig{file=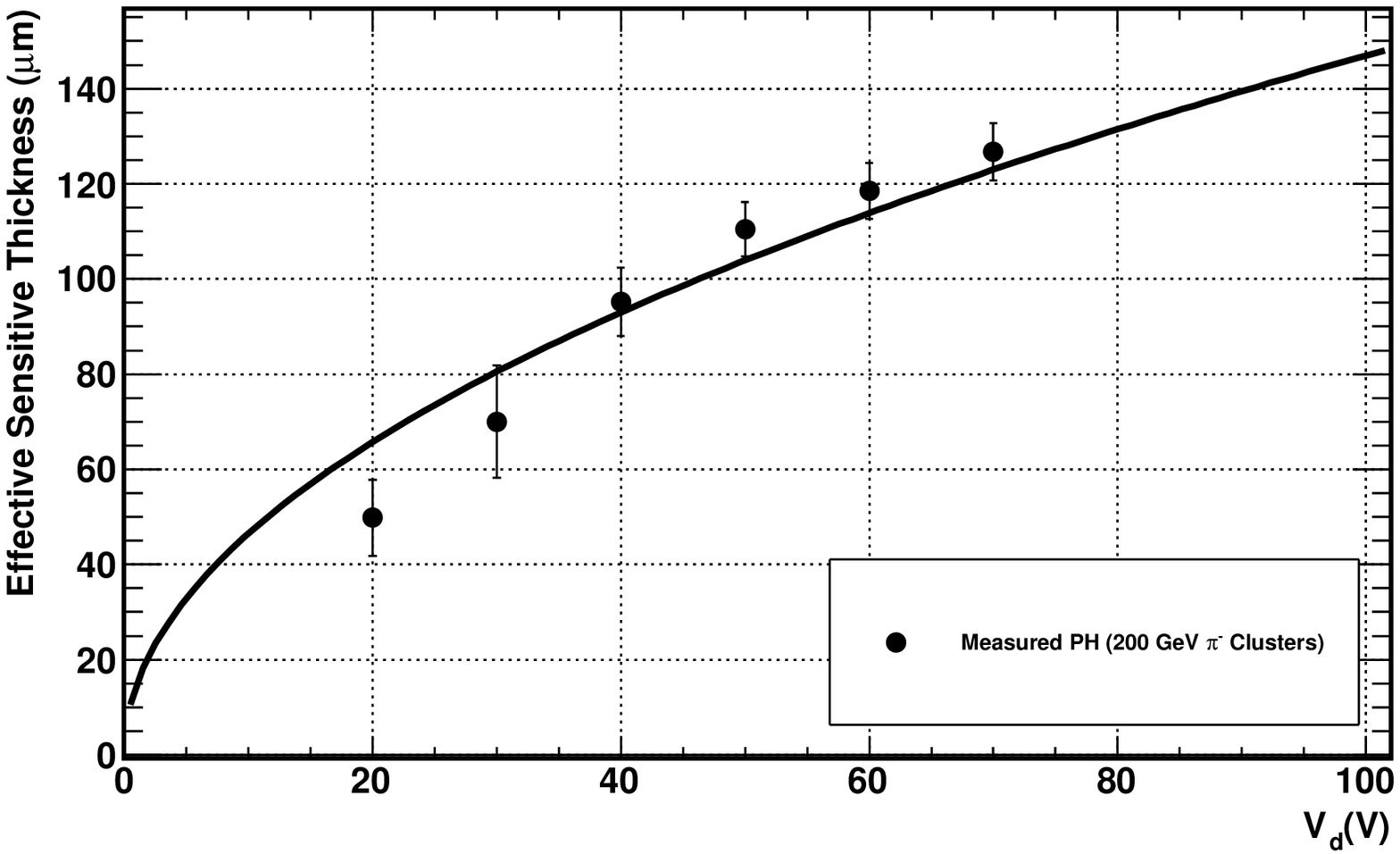,width=9.5cm} \\
\end{tabular}
\end{center}
\caption{Evolution of signal with depletion voltage: (upper) Most probable value of the charge 
signal collected in a cluster (normalised to that at 50~V) as a function of $V_d$ for 
200 GeV $\pi^-$ (filled circles) and 980~nm (filled squares) and 1060~nm (open squares) laser pulses. 
The scaling of the collected charge is compared to that of the number of clusters for $^{55}$Fe and 
the depletion thickness from the $C-V$ measurements (continuous line). (Lower) Effective sensitive 
thickness from the 200 GeV $\pi^-$ data (point with error bars) and the fit according to Eq.~(1).}
\label{fig:charge}
\end{figure}

The measurement of the $C-V$ characteristics allows us to estimate the thickness of the 
depleted volume, provided the current in the substrate is not too large. The pixel guard-ring 
grid is used as electrode and the detector area is assimilated to a single large diode. 
For this measurement the guard-ring is kept at ground potential and the depletion 
voltage applied to the metallised back-side. Since the determination of the effective 
area used to derive the capacitance is affected by a large uncertainty, we use the 
$C-V$ measurement only for relative measurements. The solid line in the top panel of 
Figure~\ref{fig:charge} shows the depletion depth, relative to that at 50~V, inferred from 
the measured capacitance as a function of the applied substrate bias. When this measurement 
is repeated by applying $V_d$ to the biasing $n$-type guard-ring, the result agrees. 

\begin{figure}
\begin{center}
\epsfig{file=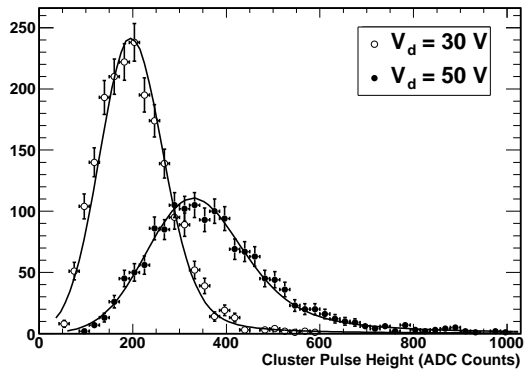,width=8.0cm} 
\end{center}
\caption{Distribution of the pulse height signal for clusters associated to a reconstructed pion 
track at $V_{d}$ = 30~V (open dots) and 50~V (filled dots). The fitted curves are convolutions 
of a Landau function with a Gaussian noise term.} 
\label{fig:PH}
\end{figure}

The results from the C-V characteristics are compared to the evolution of the amount of charge 
collected as a function of the depletion voltage using  5.9~keV X-rays, 980~nm and 1060~nm wave-length 
laser pulses, and  200~GeV pions, which are virtually un-attenuated through the full detector thickness. 
Figure~\ref{fig:PH} shows the pulse height measured for clusters associated to a reconstructed track 
in the data collected with 200~GeV $\pi^-$.
The evolution of the collected signal with the applied depletion voltage provides us with information on 
the thickness over which charge is collected in the pixel. If the charge is generated uniformly along 
the entire detector thickness, as in the case of the 1060~nm laser and the pions, we expect the measured 
signal to increase proportionally to $\sqrt{V_d}$, i.e.\ as the depleted 
thickness. On the contrary, if the penetration depth is smaller, the pulse height should reach a 
saturation level when the depleted thickness exceeds the penetration depth. Another estimator of 
the sensitive thickness is the rate of $^{55}$Fe X-rays clusters detected as a function of $V_d$, 
which we expect to scale proportionally to the thickness of the sensitive region. We determine the 
most probable value of the cluster pulse height by performing a fit with a Gaussian curve to the 
laser data and with a Landau function folded with a Gaussian noise term for the pion data. Results are 
summarised in Figure~\ref{fig:charge}, which shows the most probable value of the measured cluster pulse 
height for the 980~nm and 1060~nm laser and for the 200~GeV pions and the rate of $^{55}$Fe clusters, 
normalised to the values obtained for the reference value of $V_d$ = 50~V, as a function of the depletion 
voltage. 
We observe that the recorded pulse height increases, as expected, with the depletion voltage for both the 
pions and the 1060~nm laser. The scaling of the pulse height for clusters from laser pulses and pions is 
in remarkable agreement with that obtained from the $C-V$ characteristics curve, which scales as $V_d$. 
The relative rate of $^{55}$Fe clusters also agrees with this scaling. Instead, for the 980~nm laser data 
we observe a saturation starting at $V_d \simeq$60~V. 
Using the absorption coefficient $a$ in Si from~\cite{abssi}, we compute the Si thickness in which 85~\% 
of the charge carriers are generated to be (130$\pm$25)~$\mu$m, where the quoted uncertainty represents 
the effect of a 15~\% change in the value of the Si absorption constant, consistent with observations 
in~\cite{abssi}. We estimate the effective depleted thickness as a function of the depletion voltage, 
assuming that the measured signal scales as:
\begin{equation}
{\mathrm PH} = k \sqrt{2 \epsilon \rho \mu V_{d}}
\end{equation}
and perform a 1-par $\chi^2$ fit to extract the effective resistivity of the bulk Si from the pion 
data (see Figure~\ref{fig:charge}). 
We obtain $\rho_{eff}$ = (753 $\pm$ 41)~$\Omega$$\cdot$cm. This value would correspond to the real resistivity 
of the Si if all the collected charge were generated in the depleted layer, while we cannot exclude 
that the undepleted bulk also contributes to the detected signal. In any case, the fitted value is in 
reasonable agreement with the nominal resistivity of the SOI wafers of 700~$\Omega$$\cdot$cm, declared by 
the manufacturer. Using the fitted value of the resistivity we obtain an effective sensitive thickness 
of 115~$\mu$m at $V_d$=60~V.

\begin{figure}
\begin{center}
\begin{tabular}{cc}
\epsfig{file=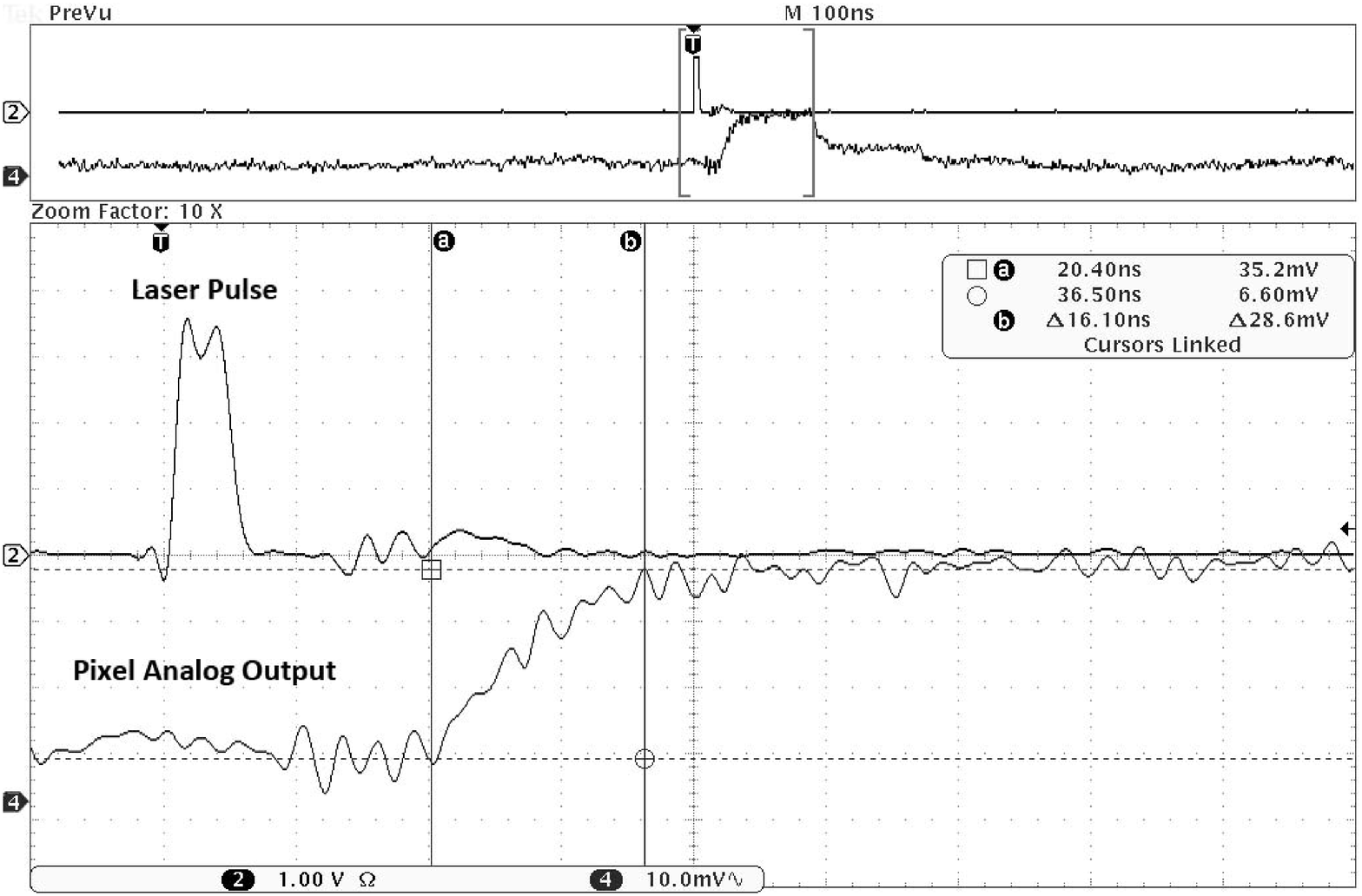,width=6.5cm} &
\epsfig{file=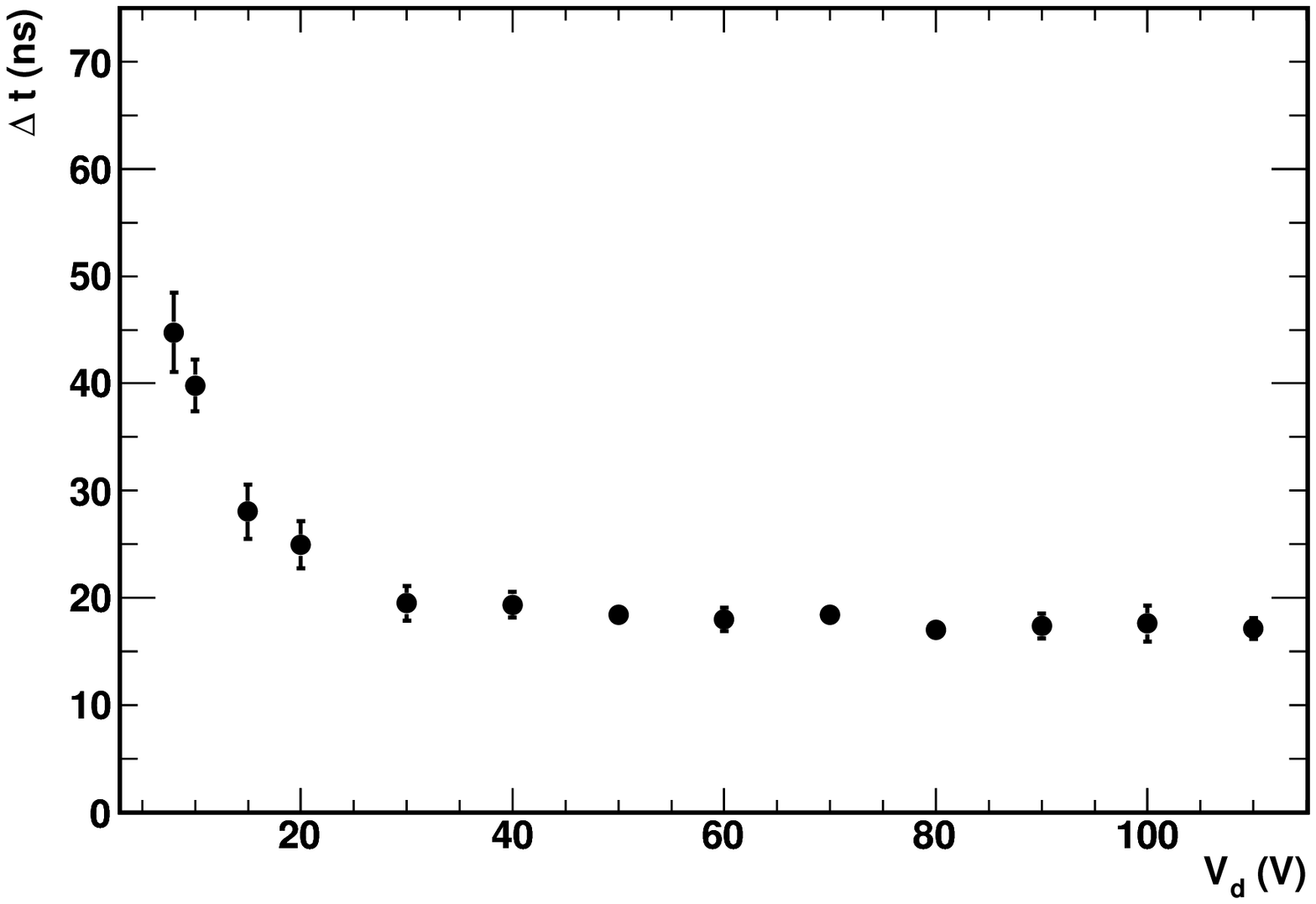,width=7.0cm} \\
\end{tabular}
\end{center}
\caption{Charge collection time measured with 980~nm laser front-illumination. (Left) Oscilloscope 
traces of a measurement. The upper trace shows the pulse driving the laser and the lower trace the 
analog output of the pixel hit by the laser beam. The two vertical bars indicating the rise time 
of the signal on the pixel are 13.4~ns apart. (Right) Average signal rise time as a function of 
the depletion voltage.}
\label{fig:time}
\end{figure}

Finally, we study the charge carrier collection time from the evolution of the signal on a pixel hit 
by a highly collimated laser pulse. To perform this measurement, a 2~ns-long pulse of a 980~nm laser is 
synchronised to shine the detector at the start of the 80~ns-long read-out window 
of the pixel onto which the pulse is focused. The analog signal  is sampled at the pixel output using a 1~GHz 
digital oscilloscope. The time elapsing between the end of the laser pulse and the time at which the 
analog signal reaches a plateau is measured at various depletion voltages on the oscilloscope. We observe 
a decrease of this time interval with increasing values of $V_d$ until it stabilises at $\simeq$17~ns 
for depletion voltages in excess of 40~V. This value does not necessarily reflect the real collection 
time since it could be due to the settling time of the read-out electronics, but demonstrates the 
effect of increasing electric field on the charge carriers. Figure~\ref{fig:time} shows the observed 
analog signal trace and the average time interval obtained from a series of ten subsequent measurements 
as a function of $V_d$, for sensor front illumination.  

\subsection{Signal sharing among pixels}

\begin{figure}
\begin{center}
\epsfig{file=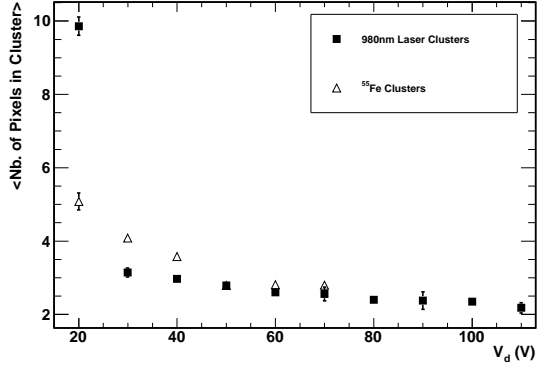,width=8.0cm} 
\end{center}
\caption{Pixel multiplicity in signal clusters as a function of $V_d$ for 980~nm 
laser pulses (filled squares) and 5.9~keV X-rays (open triangles).} 
\label{fig:npixels}
\end{figure}
Figure~\ref{fig:npixels} shows the average multiplicity of the pixels in a cluster for 980~nm 
laser pulses and 5.9~keV X-rays as a function of the applied depletion voltage, $V_d$. 
We observe that, although the charge spread decreases with $V_d$, the signal remains distributed 
over multiple pixels also for large voltages.
\begin{figure}
\begin{center}
\begin{tabular}{cc}
\epsfig{file=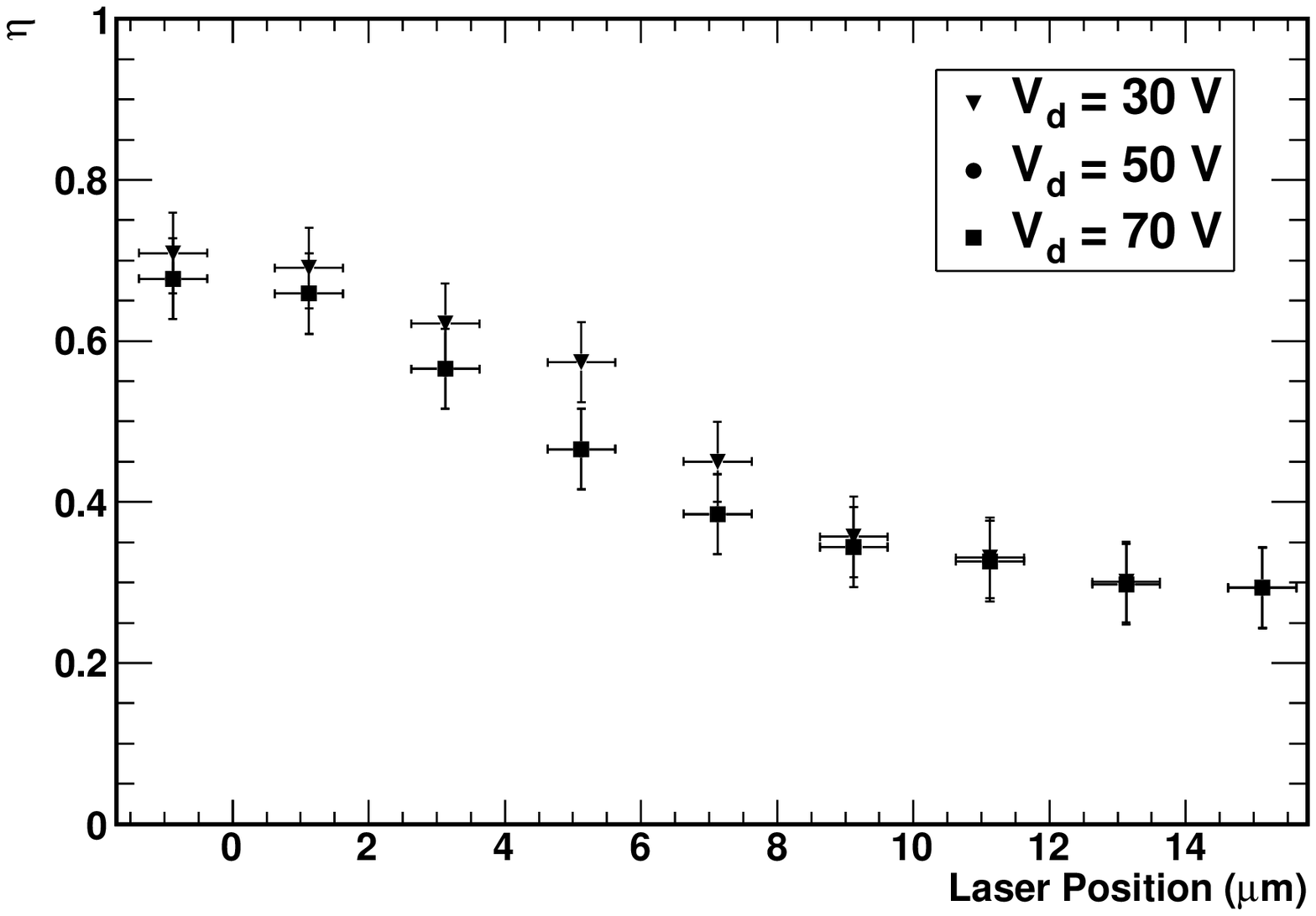,width=7.0cm} &
\epsfig{file=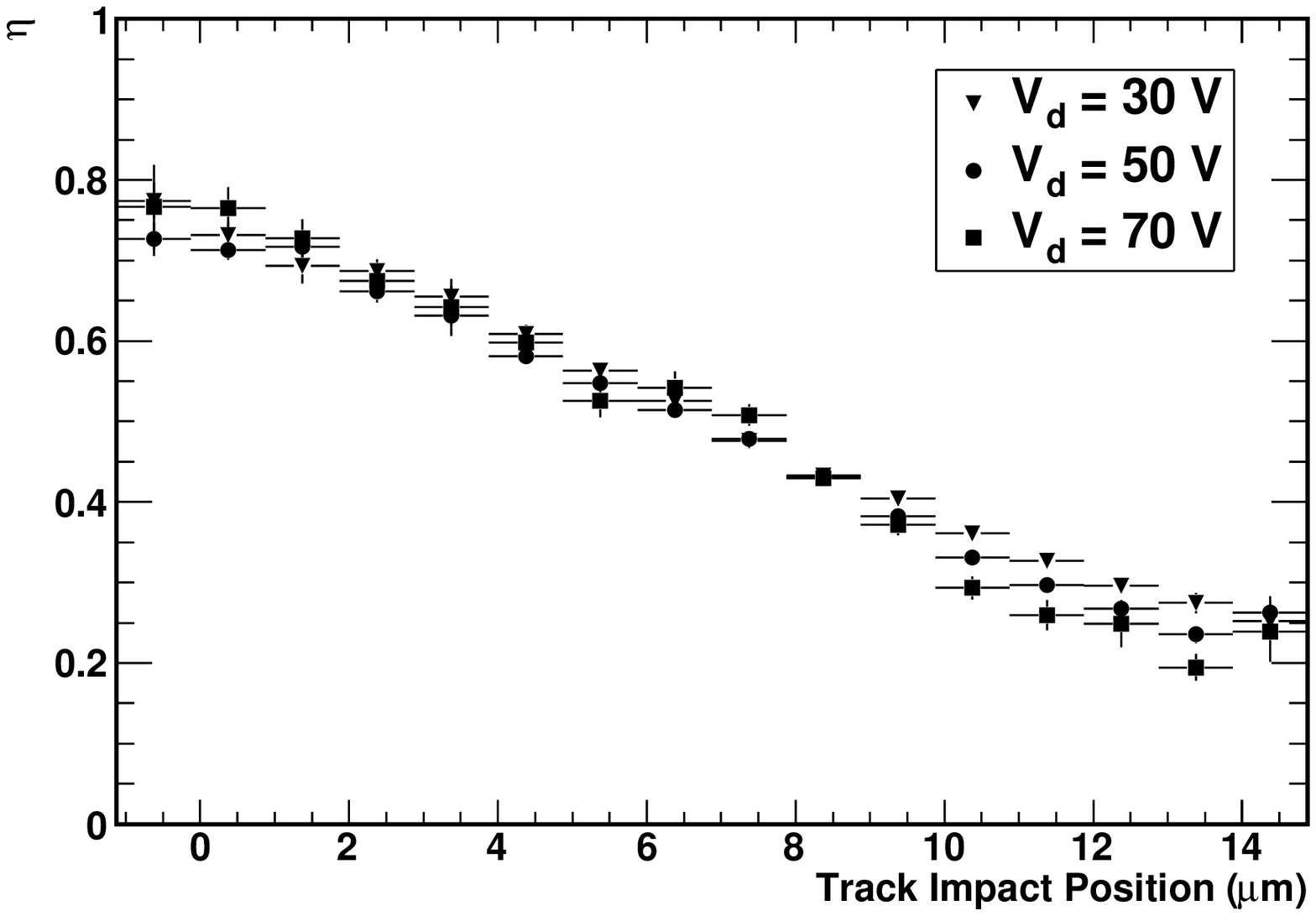,width=7.0cm} \\
\end{tabular}
\end{center}
\caption{$\eta$ distribution at different values of $V_d$ for (left) 980~nm laser 
pulses and (right) 200~GeV $\pi$.}
\label{fig:eta}
\end{figure}
Therefore, we study the charge sharing on neighbouring pixels as a function of the position of 
generation of carriers in the detector. This is usually characterised by the $\eta$ distribution 
defined as $\eta = \frac{PH_L}{PH_L+PH_R}$, where $PH_L$ and $PH_R$ denote the pulse height 
on the left and the right, respectively. Originally introduced in the characterisation of strip 
detectors~\cite{eta}, the definition of $\eta$ is extended to pixel detectors by summing the pixel 
charge over each column. The $\eta$ distributions obtained for 980~nm laser pulses and 200~GeV pions 
are shown in Figure~\ref{fig:eta}. We observe a small variation of the $\eta$ distribution with 
$V_d$ using the 980~nm laser and no significant variation with energetic pions. This result seems to 
indicate that the charge sharing among neighbouring pixels is not dominated by the charge carrier 
cloud size and that, instead, the pixel capacitive coupling plays a significant role in determining 
the observed signal distribution.

\section{Detector tracking performance}

The detector tracking performance is studied using the the 200~GeV pion data and is 
characterised in terms of the detection efficiency, the single point resolution and 
the response to inclined tracks. 
 
\subsection{Detection efficiency}

\begin{figure}
\begin{center}
\epsfig{file=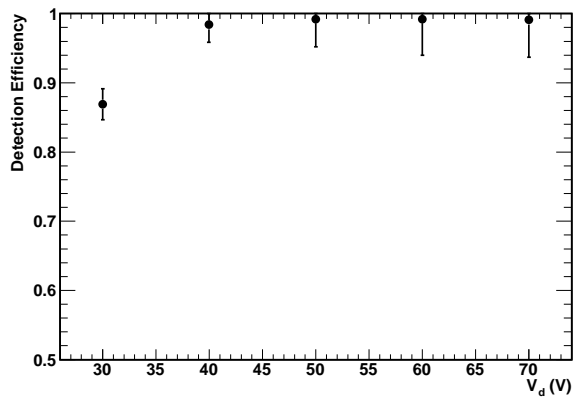,width=8.5cm} 
\end{center}
\caption{Detection efficiency as a function of the depletion voltage $V_d$.}
\label{fig:eff}
\end{figure}

The detection efficiency is measured as the fraction of tracks reconstructed on the first layer
of the doublet and on the singlet layer of the telescope, according to the criteria given above, 
which have a reconstructed hit on the second doublet layer within 10~$\mu$m from the position of 
the track extrapolation. Figure~\ref{fig:eff} shows the detection efficiency as a function of $V_d$. 
We observe a sharp rise, due to the increase of the signal S/N with $V_d$ from the increase 
of the sensitive thickness, up to 40~V. For larger voltage values, the detection efficiency 
stabilises at the value of 0.99$^{+0.01}_{-0.03}$, where the uncertainty is statistical.

\subsection{Spatial resolution}

\begin{figure}
\begin{center}
\begin{tabular}{cc}
\epsfig{file=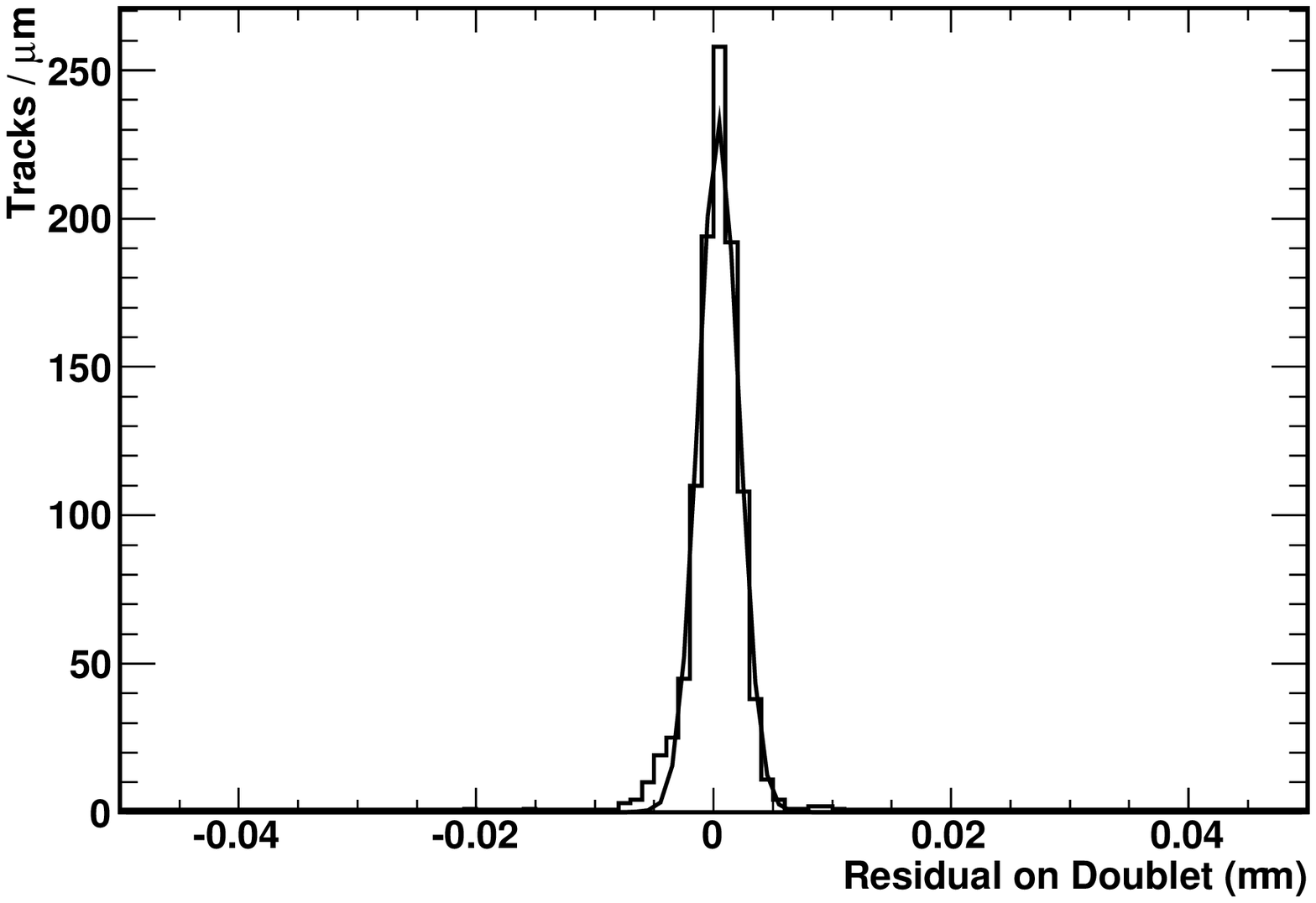,width=7.0cm} &
\epsfig{file=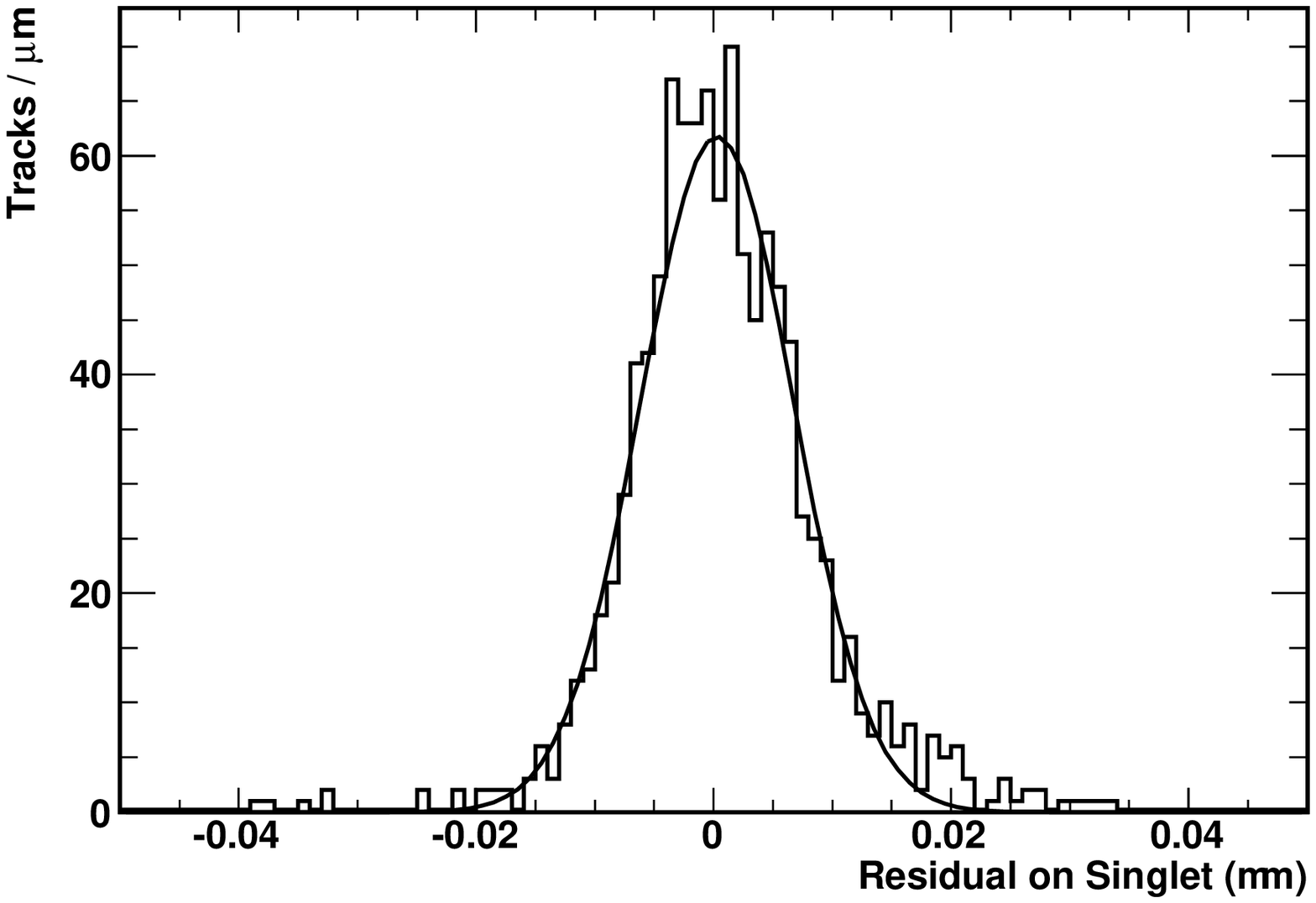,width=7.0cm} \\
\end{tabular}
\end{center}
\caption{Residual distribution for pion tracks extrapolated on (left) the second plane of the doublet 
and (right) the plane of the singlet at $V_d$ = 50~V. The fitted Gaussian curves have widths of 
(1.48$\pm$0.06)~$\mu$m and (5.83$\pm$0.23)~$\mu$m, respectively.}
\label{fig:res}
\end{figure}

The single point resolution is extracted from the Gaussian width of the residual distributions 
for reconstructed 200 GeV pion tracks. The residuals are defined as the differences 
between the reconstructed coordinate of the point of impact on the DUT and that extrapolated 
from the track reconstructed on the other two detector planes. Figure~\ref{fig:res} shows the residual 
distributions obtained using alternatively the second plane of the doublet and the singlet plane as DUTs. 
The measured width of the residual distribution is the quadratic sum of the single point resolution 
of the DUT and the track extrapolation accuracy. This, in turn, depends on the single point 
resolution of the other planes. Since the three planes are all equipped with the same detectors, 
operating at the same depletion voltage, we assume they all have the same point resolution. 
We use a {\tt Geant~4}-based simulation to relate this single point resolution to the measured residual 
width. This includes the effects of multiple scattering, though they are small at the pion energy used 
in our tests, and hadronic interactions. Residuals are computed using both the second layer of the doublet, 
where the track extrapolation accuracy is higher, and the singlet, where the sensitivity to the point 
resolution is higher, as DUT. In computing the extrapolation resolution on the DUT,we assign an 
uncertainty of $\pm$0.5~mm on the longitudinal position of the doublet and $\pm$1.0~mm on that of the 
singlet layer. Results are summarised in Table~\ref{tab:res} for various values of $V_d$. For $V_d \ge$50~V, 
the point resolution is measured to be (1.12$\pm$0.03)~$\mu$m, where the quoted uncertainty includes the 
statistical and the systematic uncertainties on the DUT longitudinal position.
\begin{table}
\caption{Measured single point resolution for different $V_d$ values, the quoted uncertainties include the
systematics from the DUT longitudinal position.} 
\begin{center}
\begin{tabular}{|l|c|c|}
\hline
$V_d$ (V)     & \multicolumn{2}{|c|}{$\sigma_{\mathrm{point}}$ ($\mu$m)} \\
              & Doublet & Singlet \\
\hline \hline
30      & 1.43$\pm$0.06             &  1.30$\pm$0.06            \\
40      & 1.18$\pm$0.03             &  1.12$\pm$0.05            \\
50      & 1.14$\pm$0.03             &  1.02$\pm$0.06            \\
60      & 1.15$\pm$0.07             &  1.10$\pm$0.08            \\
70      & 1.08$\pm$0.06             &  1.05$\pm$0.07            \\
\hline
\end{tabular}
\end{center}
\label{tab:res}
\end{table}

\subsection{Response to inclined tracks}

\begin{figure}
\begin{center}
\epsfig{file=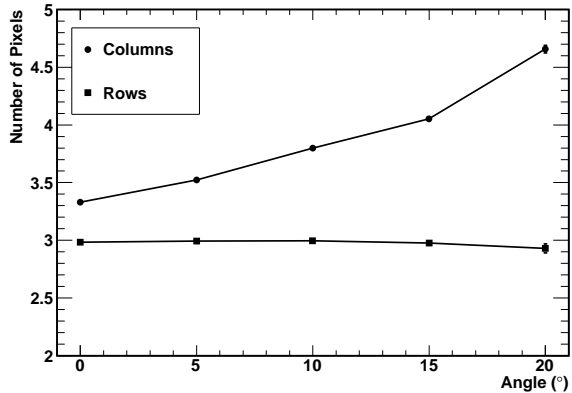,width=8.5cm} 
\end{center}
\caption{Average pixel multiplicity along rows and columns for clusters associated 
to a pion track as a function of the detector angle for $V_d$ = 60~V.}
\label{fig:angle}
\end{figure}

Finally, we study the cluster shape as a function of the angle between the incident pion 
track and the normal to the detector plane on the singlet plane of the telescope. The 
change of the cluster shape as a function of the angle of incidence is important for assessing
the detector response to forward tracks in a vertex tracker with a barrel layout and to 
low momentum background particles~\cite{Maczewski:2009}. In the beam test the angle of incidence 
is varied remotely by operating the rotation stage between 0$^{\circ}$ and 20$^{\circ}$. We take 
200~GeV $\pi^-$ tracks which have associated hits on all three planes and characterise the cluster 
shape as the number of pixels along rows and columns in the cluster associated to the pion track on 
the third plane. Since the rotation axis is parallel to the pixel columns we expect to observe an 
increase of the multiplicity along the column and a constant average number of pixels along the rows. 
Figure~\ref{fig:angle} shows the average number of pixels along columns and rows in clusters on the 
third plane as a function of the detector angle. 

\section{Conclusions}

A prototype pixel chip in OKI 0.20~$\mu$m SOI technology with 13.75~$\mu$m pixel 
pitch has been designed, processed and characterised in the laboratory with X-rays 
and short IR laser pulses and on a CERN SPS beam-line with 200~GeV $\pi^-$.  
The use of a buried $p$-well implanted beneath the buried oxide avoids the 
back-gating effect and the tested chip demonstrates for the first time large 
charge collection, high efficiency and micron resolution in a beam test. 
The chip can be properly operated to voltages $\ge$70~V, corresponding to an 
estimated thickness of the depleted region $\ge$125~$\mu$m. We study the 
evolution of the collection time for the charge generated by a 980~nm laser 
in front illumination as a function of $V_d$ and measure collection times below 
20 ns for $V_d \ge$ 50~V. The hit reconstruction efficiency for high energy pions 
has been measured to be $\ge$0.99 and the single point resolution 
(1.12$\pm$0.03)~$\mu$m, for $V_d \ge$ 50~V. These results are very encouraging 
for the further development of pixel sensors in SOI technology for applications 
demanding particle reconstruction with high spatial resolution and efficiency.  

\vspace*{-0.1cm}

\section*{Acknowledgements}

\vspace*{-0.1cm}

This work was supported by the Director, Office of Science, of the 
U.S. Department of Energy under Contract No.DE-AC02-05CH11231 and 
by INFN, Italy. We are grateful to Y.~Arai for his effective 
collaboration in the SOIPIX activities. 
We acknowledge support from CERN for the beam test at the SPS.
We are indebted to M.~Tessaro for the detector doublet mounting, 
T.S.~Kim and A.~Onnela for the setup at the beam test and to A.~Behrens 
and A.~Froton for the detector alignment on the H4 beam-line. 
We are also thankful to N.~Alster and N.~Pozzobon for their 
contribution to the data taking at CERN.

\end{document}